# FORENSICS ACQUISITION OF IMVU: A CASE STUDY

Robert van Voorst
National Police of the Netherlands
Rotterdam, Netherlands
rvanvoorst@politie-rijnmond.nl
M-Tahar Kechadi, Nhien-An Le-Khac
University College Dublin
Dublin 4, Ireland
{tahar.kechadi,an.lekhac}@ucd.ie

**ABSTRACT**

There are many applications available for personal computers and mobile devices that facilitate users in meeting potential partners. There is, however, a risk associated with the level of anonymity on using instant message applications, because there exists the potential for predators to attract and lure vulnerable users. Today Instant Messaging within a Virtual Universe (IMVU) combines custom avatars, chat or instant message (IM), community, content creation, commerce, and anonymity. IMVU is also being exploited by criminals to commit a wide variety of offenses. However, there are very few researches on digital forensic acquisition of IMVU applications. In this paper, we discuss first of all on challenges of IMVU forensics. We present a forensic acquisition of an IMVU 3D application as a case study. We also describe and analyse our experiments with this application.

## 1. INTRODUCTION

Instant messaging (IM) is one of the most popular digital communication technologies and is widely used today. In fact, IM belongs to the category of Internet based communication services. Traditionally the IM was designed to transfer text messages only, but now it has been enhanced with additional features such as voice/video message, file transfer, etc. Moreover, IM can be accessible through cell phones or other mobile devices, as well as on a computer. It is estimated that there are several millions of IM users who use IM for various purposes. There are numerous IM applications available for download today [1].

Besides, IM applications (apps) also allow users to create a detailed personal profile including name, email address, age, home address, phone number, school and hobbies. If users do not verify carefully during the sign-up process, they could reveal more information than they should. Easy accessible profiles can allow anyone to contact IM users. Indeed, some IM apps offer users the option of joining in chat with strangers. As a consequence, using of IM apps can encourage gossiping and bullying. Children could receive pornographic "spam" through IM apps or become victim of Sexual grooming of children [2]. In the recent years, police investigators have noticed that IM apps have been directly or indirectly involved in criminal activities. Due to their popularity, IM apps have the potential of being a rich source of evidential value in all kinds of criminal investigations. Indeed, Europol has identified the threat of misused IM communications by criminals to facilitate their illegal activities due to the fact that it is harder to monitor or to regulate these services [3].

On the other hand, a virtual universe is a computer-based simulated environment. In a 3D virtual universe, the users can take the form of 3D avatars (virtual character) visible to others. Today, Instant Messaging within a Virtual Universe (IMVU) combines custom avatars, chat/IM, community, content creation, commerce, and anonymity [4].

IMVU contains its own economy with a currency system based on IMVU "credits" and "promotional credits". Credits can be purchased online using actual currency either directly from IMVU or from third party re-sellers. IMVU members use the credits to purchase virtual items like fashion pieces (hair, clothes, skins, and accessories), pets, and 3D scenes such as homes, clubs and open landscapes. Except 'VIP users', everyone has "Guest_" in front of his/her avatar name. To become a VIP you need to buy a subscription. There are benefits for VIPs such as extra credits, block unwanted avatar actions, unlock VIP-only avatar actions and whisper privately in chats. It is also possible to buy an 'Access Pass'. The Access Pass is available for those 18 years old and older. It grants access to more risqué clothing and slightly more intimate furniture and poses. It also allows chatting to adults only.

IMVU was also mentioned as a common Virtual World addressing 'Crime and Policing in Virtual Worlds' (2010) by Marc Goodman [5]. He concludes that criminals can exploit this new technology to commit a wide variety of offenses. Almost any types of crime in the real world can also be found in virtual spaces from child abuse to terrorist attacks. Besides, the presence of child pornography in chat has also been denounced in 3D social networks like IMVU. Indeed, the emerging of 3D social networks is rapidly tainted with child pornography as stated in [6].

In fact, there are many tools available for collecting artefacts, some of them specific for chat/IM such as Paraben chat examiner [7] or Belkasoft Evidence Centre 2015 [8]. However, at the time of writing this paper, none of them supports the investigation of IMVU apps. Despite that IMVU exists a while, has millions of users worldwide and it is associated with child abuse (material), to the best of our knowledge there is any technique or software tool for forensic acquisition and analysis of IMVU artefacts. Therefore, the objective of this paper is to investigate whether it is possible to forensically acquire and analyse artefacts on a computer after the installation and use of the IMVU application.

The rest of this paper is structured as follows: Section 2 shows the background of this research including related work in this domain. We present IMVU apps and forensic process in Section 3. We describe our case study in Section 4 and discuss on experiment results in Section 5 and 6. We conclude and discuss on future work in Section 7.

## 2. RELATED WORK

In fact, we have tried to find in literature about this specific subject, but to the best of our knowledge, there is no published research regarding forensic artefacts of IMVU. Hence, we present some literature survey on the forensic acquisition and analysis of IM applications in general. A more traditional investigation method of IM has been described in [9]. This handbook has covered basic instructions on the procedures to gather digital information from IM such as: the screen or user name (victim and suspect), the IM service being used and version of the software, the content (witness account of contact or activity), the date and time the message was received/viewed, the dates and times of previous contacts, any logging or printouts of communications saved by the victim, applicable passwords. Besides, when investigating crimes involving IM services, investigators normally extract the following information from a chat room: (i) Name of the chat room; (ii) Web address of the chat room; (iii) Screen or user name (victim and suspect); (iv) Chat software being used, and version of the software; (v) Content (witness account of contact or activity); (vi) Date and time the communication took place; (vii) Dates and times of previous sessions where similar activity took place; (viii) Logging or printouts of communications saved by the victim and (ix) Applicable passwords.

Kiley et al [10] conducted some research on IM forensic artefacts and they also pointed out that IM is being exploited by criminals due to its popularity and privacy features. As mentioned in their paper, one of the challenges for forensic examiners is the web-based instant messaging when they don't write to registry keys or files. Apart from the typical conversation log showing on screen, and possible chat log saved on the logical drive, evidence of conversation could also be found in page files and unallocated hard disk due to the volatile nature of the data. Investigators should, therefore, look for remnants of whole or partial conversations that may be dumped to page files and unallocated space on the hard disk. One of the conclusions from this work is that forensic evidence is recoverable after these programs have been used, but investigators must know certain elements of the conversations in order to perform string searches. Even so, time consuming sector-by-sector searches is huge to uncover all the potential evidence.

Lun [11] describes the popularity of IMs and the challenges by extracting information from these IMs by Law Enforcement. He has reviewed techniques and forensic tools designed by different developers to extract relevant information. He also posed the question; 'are the current forensic techniques and tools adequate?' and presents different techniques that might apply on extracting information under different scenarios. He also listed useful sources of information such as the Windows registry, physical memory (RAM), hibernation file and page file. An off the shelf forensic tool Belkasoft Evidence Center was used to extract evidence from the MSN live messenger environment. His finding indicated that the forensic tool was able to aid the forensic examiner to extract digital evidence from instant messenger, but it was not comprehensive enough at that forensic examiners could not rely on one tool during an investigation.

Dickson [12] describes the examination of AOL Instant Messenger (AIM) 5.5. This research focuses on the contact information (identification) and the ability to trace back from a known suspect's computer system to identify contact with his alleged victims. This is specifically interesting when an allegation of grooming is made. According to the author, one of the first things an examiner should do is to determine the suspect's computer system that was used to contact the informant's account in the first place. The author also states that this is a strong evidential link between victim and suspect and it also clearly demonstrates that the suspect was responsible for that contact rather than a third party who may have taken control of the suspect's chat account. The author moreover illustrates some methods of proving a link where the suspect and victim have been in contact on the Yahoo Messenger chat service.

Husain et al. [13] describe the Forensic Analysis of Instant Messaging on Smart Phones. They studied and reported the forensic analysis of three different IMs: AIM, Yahoo! Messenger and Google Talk on Apple iPhone. Their results have shown that various useful artefacts, related to IMs, can be recovered, including username, password, buddy list, last log-in time, and conversation timestamp as well as conversation details.

In [14], authors observed that the relevant research on the topic of evidence collection from IM services was limited. Authors presented the forensic acquisition and analysis of WhatsApp, Viber, Skype and Tango IMs and VoIPs for both iOS and Android platforms and authors tried to answer on how evidence can be collected when IM communications are used. Authors also referred to the fact that Europol has identified the threat of misused IM communications by criminals to facilitate their illegal activities due to the fact that it is harder to monitor or regulate these services. The objective was to identify the artefacts stored by each IM application in the file system of every seized device. This work provides useful information that assists us in developing our case study in this paper. More precisely the authors have provided a very useful taxonomy of target artefacts.

## 3. IMVU AND FORENSIC PROCESS

In this section, we describe our forensic process for investigating IMVU apps. The objective of this paper is to forensically examine significant artefacts present on a computer after the installation and use of the IMVU apps. Therefore, the approach chosen is empirical or experimental research and differential forensic analysis. While there is no available technique or software tool to evaluate and there is very little known about IMVU artefacts, our approach includes the collection of data during experiments and try to discover significant forensic artefacts by observing and analysing the data. In fact, with an IMVU app, apart from the typical conversation log showing on screen, and possible (log) files saved on the logical drive, evidence of conversation could also be found in page files and unallocated hard disk space, the windows registry or volatile data in the internal memory (RAM). There are two main steps in our forensic process: data collection and data analysis.

IMVU data collection: in this phase data is collected during the running time of an IMVU app. The contents of the IMVU application and data directory, unallocated disk space, the Windows registry, volatile data from the internal memory and network traffic are acquired. We attempt moreover to infer the behaviour of the IMVU application through the reverse engineering process. The objective is to trace user activities, like the history of performed activities such as instant messages sent or received (chat logs), date and time stamp of the communication, sender, receiver and password. All experiments were carried out several times to ensure that results are repeatable.

IMVU data analysis: This phase gives meaning to the data that have been collected so that the research question can be answered. Differential forensic analysis [15] is carried out by comparing the collected data (snapshot) before and after the experiments with IMVU to observe the differences between them such as report the files that have been added, deleted, renamed, and altered, the difference between Windows registry hive files. The objective of the analysis is to understand the changes and to determine if it contains any relevant information or not. A taxonomy of target artefacts is defined in order to guide and structure subsequent forensic analysis. The taxonomy used in our analysis is described in [14].

## 4. IMVU FORENSICS: A CASE STUDY

In our case study, we use computers with 4GB of RAM, 30GB hard drive space formatted with the NTFS file system and network access (using a static IP address). The testing was conducted using Windows 7 Enterprise, SP1. Besides, the monitoring tool SysTracer was installed onto the system to record file activity and changes made to the Windows registry. Another computer was setup as a man-in-the-middle proxy using MITMproxy, running on the Ubuntu 14.04.1 LTS, this computer is used to monitor and intercept network traffic as shown in Figure 1. Next, an IMVU client program, version 514.0, was installed. In order to use IMVU, firstly a user has to create an account. For this case study, we create three IMVU accounts (InspectorAlgar, LadyElizabethSmall and missycanaryyellow). A valid Avatar name should contain letters and numbers and must be between 3 and 20 characters. We conduct different conversations to create test data. The conversations were in public rooms but also limited by two participants. We created moreover an AutoIt [16] script that simulated an IMVU user. On beforehand a text file was created containing the conversation text of one of the discussion partners. A chat message was send to the IMVU application within a fixed (pre-adjustable) interval. By starting two scripts on different machines in appropriate time, we can simulate a chat conversation. The chat messages contained, the chat text and a Unix timestamp. An example from the chat messages is shown in Figure 2. In the next two sections we present and analysis the forensic results of this case study related to the file system (Section 5) and the internal memory (Section 6).

## 5. IMVU: FILE SYSTEM FORENSICS

In this section, we describe our forensic acquisition results of IMVU file systems including application file, log files, profile data and location data.

### 5.1 Application Files

IMVU application files are stored in: *C:\Users\<Username>\AppData\Roaming\IMVUClient\* folder. This path is fixed during the installation. During the installation a shortcut to *IMVUQualityAgent.exe* is added into the Windows Start-up folder as well on the user's desktop. We also found that IMVU uses Mozilla's Gecko rendering engine and the XML User Interface Language for building its user interface [17].

Besides, a configuration file *imvu.cfg* is added into the user Flash Player Trust folder: *C:\Users\%USERNAME%\AppData\Roaming\Macromedia\FlashPlayer\#Security\FlashPlayerTrust\*. Configuration files inside this *FlashPlayerTrust* folder normally contain a list of directory paths. We also observe that application shortcuts are added to IMVU folder: *C:\Users\%USERNAME%\AppData\Roaming\Microsoft\Windows\StartMenu\Programs\IMVU\*. We found moreover the IMVU application is written in Python and the compiled script files are stored in *library.zip*: *C:\Users\Robert\AppData\Roaming\IMVUClient\library.zip*. We also decompile these files using Easy Python Decompiler and to explore the IMVU application through reverse engineering.

### 5.2 Log Files

IMVU log files are stored in *C:\Users\<Username>\AppData\Roaming\IMVU\* folder including *cpp.log, IMVUQualityAgent.log, IMVULog.log,* etc. The *cpp.log* contains information related to user interface and this file is created every time the application starts. The *IMVUQualityAgent.log* file is also created every time the application starts. It contains 'application validation' information and local time stamp. Regarding to *IMVULog.log* file, from the digital forensics examination point of view, this file contains the most valuable information. In fact, this log file is generated by a roud-robin process based on session and file size. A new log file is created every time the application starts. When the size of the log file is greater than 2MB (2097152 bytes), this process successively create new files with the same pathname as the first file, but with different extensions started with ".1", ".2" and so on. If it is greater than ".6", the extension restarts from ".1", older files will be deleted. We found moreover the definition of this round-robin process in the Python script *log.pyo* from *library.zip* file mentioned in Section 2.1. Each record in IMVULog.log file contains the following attributes (Figure 3):

1. **The number of seconds since start of application:** *Floating point decimal value. The length is 6 characters minimum including the dot and 3 decimals (milliseconds rounded). If the length is less than 6, a leading 0 is added.*

2. **The thread id:** *String value, length fixed with 5 characters, padded by spaces if necessary.*

3. **File name of the source file where the logging call was issued:** *String value, length fixed with 30 characters, padded by spaces if necessary*

4. **Source line number where the logging call was issued:** *Signed integer decimal value, length fixed with 4 digits, padded by spaces if necessary.*

5. **The logging level for the message:** *String value, length fixed with 7 characters, padded by spaces if necessary. Possible values are 'DEBUG', 'INFO', 'WARNING', 'ERROR', 'CRITICAL'.*

6. **The log message**: *String value.*

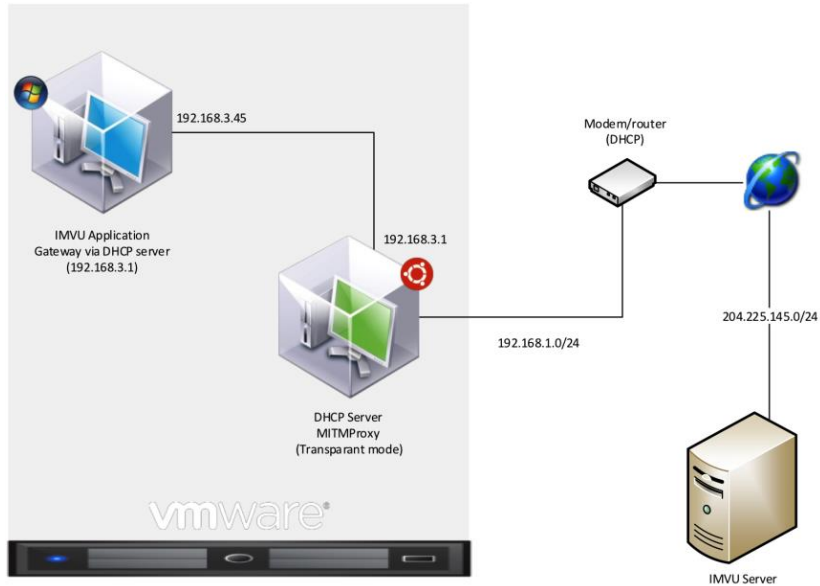

Figure 1 MITM proxy setup

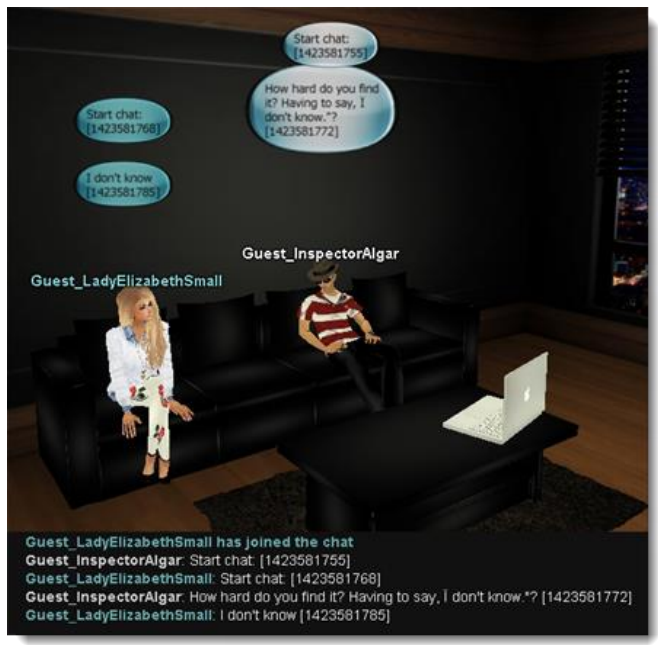

Figure 2 Chat script usage example

Moreover, we notice that the maximum length of an entire log record (attributes 1-6) is 512 characters. If the length exceeds 512 characters, it is truncated to the first 509 characters and three dots ('…') are added to the end of the record. Besides, if the length of the log message (attribute 6) is greater than a defined length this log message is truncated and three dots are also added to the end. We have also observed that the log message regarding to logging level 'ERROR', and 'INFO' consists of multiple lines, while other log messages only contain a single line.

When analysing the log files, we noticed that the IMVULog file also stores information related to text communications. To analyse an IMVU chat, it is necessary to understand the following important aspects of a chat conversation: (i) Who says what to whom; (ii) In which chat session the conversation takes place, note that an user can be present in multiple rooms (chat sessions) at the 'same' time; (iii) Timestamp of the messages.

The text of a single message appears more than once in different log records. Besides, in a chat room, it is possible to have a private conversation with just a single person in that room (at least one of users is a VIP). Whisper is a special feature available exclusively for VIPs. A VIP member can whisper to anyone and vice-versa, anyone can whisper to a VIP. Two non-VIP users however cannot use this feature to chat with each other. Moreover, the log message contains a list of several elements. The first element is the member id (*'userId'*), the second one is a chat id (*'chatId'*) and the last one contains dictionary data with keys and values. The value '0' of the key indicates that the message was sent to the participant(s) of a room. There are one or more participants in a room. The value of the key '*message*' contains the actual chat message. We also note that the value sometimes contains an asterisk followed by a chat command. Besides, if the message exceeds the maximum length, as mentioned previously, it is truncated and three dots are added to the end.

The value of the key *'userId'* contains the member id of the sender of the text message. The value of the key *'chatId'* contains a unique id for each chat session. Concurrent chat sessions in different rooms therefore have different IDs. This log record does not contain a timestamp. To determine the date and time of the message for these log records, there are two possibilities:

1. Calculate the approximate timestamp using the number of seconds since the start of the application. This number is listed at the beginning of the log record.
2. Combine it with the log records derived from the file *SessionDispatcher.pyo*.

### 5.3 User profile data

The *IMVULog* file only contains user profile data when the status is 'Running' (i.e. the application is running and the user is logged in). The following information can be retrieved from the user profile data: the *users' avatarname* associated with the IMVU account, eight (8) asterisks of the users password. Note that the real user password is always replaced by eight asterisks. It is independent on the actual length of the password. Indeed, the most comprehensive information can be retrieved from the multi line log message started with self.userInfo_ and followed by multiple lines of dictionary (pairs of <key, value>) data.

The above information is based on the data entered by the user (Account Settings) and data allocated by IMVU. Besides, objects in IMVU are identified by a unique id such as, for example, Member ID, Room(Instance) ID, Session ID, Product ID or Chat ID.

```
   ❶      ❷           ❸        ❹       ❺                        ❻
01.099 3328*          log.pyo( 509)    INFO: ------------------------------------------------
01.100 3328*          log.pyo( 510)    INFO: GenerateLogger: cwd is
                                             C:\Users\Robert\AppData\Roaming\IMVUClient
01.100 3328*          log.pyo( 511)    INFO: disableLogRotationWhileRunning: None
01.100 3328*          log.pyo( 512)    INFO: ProgramDirectory:
                                             C:\Users\Robert\AppData\Roaming\IMVUClient
01.101 3328*          log.pyo( 514)    INFO: UserDirectory:    C:\Users\Robert\AppData\Roaming\IMVU
01.101 3328*          log.pyo( 515)    INFO: Current time:     2015-02-06 11:14:51
01.101 3328*          log.pyo( 520)    INFO: Looking for logging config file at
                                             C:\Users\Robert\AppData\Roaming\IMVUClient\logging.cfg
01.102 3328*          log.pyo( 532)    INFO: No valid logging.cfg found and applied
01.102 3328*    clientapp.pyo( 713)    INFO: Starting IMVU version 516.0
01.129 3328*      network.pyo(  75)    INFO: ProxyEnable is 0, ignoring proxy settings
01.130 3328*      windows.pyo( 103)    INFO: IMVUMainThreadSerializerMessage registered as 0xc16d
01.130 3328*      windows.pyo(  31)    INFO: IMVUKillSleepMessage registered as 0xc174
01.213 3328*      windows.pyo( 154)    ERROR: Unable to set IMVU protocol handler

< truncated to improve readability >
```

Figure 3 Partial IMVU.log file

```
    imqconnection.pyo( 515)       INFO: connecting 192.168.3.45:49219 ->
204.225.145.75:443

   < truncated to improve readability >
```

Figure 4 IP address information in the IMVULog file

### 5.4 User authentication data

By investigating the Windows registry, we found that IMVU authenticate users by using username and password. This information is stored in the registry if the option 'Save Password' is selected. Indeed, the relvant password can be retrieved from the registry. The users avatar name associated with the IMVU account is also present in the *IMVULog* file. The *IMVULog* file also contains information about contacts such as 'Friends", "Buddies" or "Fans". Besides, we also found that the *buddystae.pyo* script implements the logging process.

### 5.5 Location data

There is limited information on location. This already has been described under in User profile data section, where the *'country'* key referrers to the geo-location of the IP address used and *'country_code'* referrers to the location specified by the user in the profile settings. In the *IMVULog* file the client and server IP addresses are listed. In the example shown in Figure 4, the client's internal Local Area Network (LAN) private network address (192.168.3.45) is shown. When the system is directly connected to the Internet and receives an external Wide Area Network (WAN) public IP address, this IP address is listed in the log file.

### 6. IMVU INTERNAL MEMORY FORENSICS

In this test, authors have explored the Python source code to understand how log records are buffered in memory. Whenever a record is added into the buffer, a check is made to see whether the buffer should be flushed or not. In fact, the buffer is periodically flushed whenever the buffer is full, or when an event of a certain severity happened. After making a memory dump of a running system, authors were able to retrieve log record data as previously described above. Besides, because a record consists of isolated fragments, it is difficult to link the messages to a date and time, a chat ID or the participants. We also notice that it was not possible to create a regular expression that fully matched all desired log records in EnCase because in Encase (version 6.19.7) a maximum limit of 255 characters that a single GREP expression could match is noted. Note that the maximum length of an entire log record (*the log record attributes 1-6*) is 512 characters. Therefore, there are fewer records found in EnCase than with the use of the in-house Python script. IMVU also creates a SQLite database in memory. Authors were able to modify the source code and save this database to hard disk. The database consists of only a single table and contains information about the HTTP cached files.

## 6. CONCLUSION

In this paper, we present a forensic acquisition process of IMVU app. IM is one of the most popular digital communication technologies and widely used, but also being exploited by criminals due to its popularity and privacy features. So it is important that we are able to collect artefacts of IM apps. While most IM apps use a well-known formatted history file (also known as IM log or chat log) and there are tools available for digital evidence collection, this is not the case with IMVU. IMVU has a set of fixed number log files that contain not only chat messages, but also identifiable data. IMVU app also deletes old log files when the number of log files reaches a limit number. Forensic artefacts can also be found in user profile data, authenticated data and location data. Indeed, we can also retrieve IMVU artifact from the internal memory by explore its SQLite databases.

We are working on the acquisition and analysis more forensic artefacts of an IMVU app such as registry information, network traffic data and unallocated disk space.